# A Multicore Processor based Real-Time System for Automobile management application.

[centered and aligned to line (T*, S-14*, B*)]

**Vaidehi. M[1], Dr. T. R. Gopalakrishnan Nair [2]**

[Authors Names]

[1]Research Associate, Research and Industry Incubation Centre, Dayananda Sagar Institutions, Bangalore.email: vaidehidm@yahoo.co.in

[2] Director, Research and Industry Incubation Centre, Dayananda Sagar Institutions, Bangalore. email: trgnair@ieee.org

Dr. T. R. Gopalakrishnan Nair
Director,
Research and Industry Incubation Centre,
Dayananda Sagar Institutions, Bangalore-560078, India
email: trgnair@ieee.org

**ABSTRACT**

In this paper we propose an Intelligent Management System which is capable of managing the automobile functions using the rigorous real-time principles and a multicore processor in order to realize higher efficiency and safety for the vehicle. It depicts how various automobile functionalities can be fine grained and treated to fit in real time concepts. It also shows how the modern multicore processors can be of good use in organizing vast amounts of correlated functions to be executed in real-time with excellent time commitments. The modeling of the automobile tasks with real time commitments, organizing appropriate scheduling for various real time tasks and the usage of a multicore processor enables the system to realize higher efficiency and offer better safety levels to the vehicle. The industry available real time operating system is used for scheduling various tasks and jobs on the multicore processor.

# A Multicore Processor based Real-Time System for Automobile management application.

## 1. Introduction

Requirement of high speed processors is in top of the line in real time systems, as the cost of high-speed processors has come down this has paved way for solving real time system demands effectively. Today multicore systems are reviewed in real time embedded system.

One of the most prospectus systems where multicore can be applied is in the domain of automobiles and transport management systems. Here we discuss the application of quad core processor applied to manage a standard luxury car in order to enhance its features in performance, safety, security and entertainment, multi fold.

The real time performance of multicore systems is slued with appropriate scheduling algorithms and task segregation schemes in order to have better control over the task [1]. Domain partitioning and bounded multi-processing schemes are applied .Simulation studies shows that the approach could be termed to perfectly fit a selected automobile for best success. The speed race is over. Faced with the growing energy consumption and excessive operating temperatures caused by high CPU clock speeds, microprocessor vendors have adopted a new approach to boosting system performance integrating multiple independent processor cores on a single chip.

Intel, for example, has proclaimed that all of its new CPUs will use multi-core architectures and has recently produced a roadmap that details processors based on two, four, and eight cores. Until now, multi-core processors for the desktop and server markets have governed the lion's share of media attention. But multi-core is also taking root in the embedded industry, with the introduction of processors such as the dual-core Freescale MPC8641D, the dual-core Broadcom BCM1255, the quad-core Broadcom BCM1455, and the dual-core PMC-Sierra RM9000x2.

Multi-core processors like these are poised to bring new levels of performance and scalability to networking equipment, control systems, videogame platforms, and a host of other embedded applications. Compared to conventional uniprocessor chips, multi-core processors deliver significantly greater compute power through concurrency, offer greater system density, and run at lower clock speeds, thereby reducing thermal dissipation and power consumption issues [2].

There is one problem, however most software designers and engineers have little or no expertise in the programming models and techniques used for multi-core chips. Instead of relying on increasing clock speeds to achieve greater performance, they must now learn how to achieve the highest possible utilization of every available core.

For instance, the challenge of managing shared resources in a multi-core chip. In most cases, the cores have separate level 1 caches, but share a level 2 cache, memory subsystem, interrupt subsystem, and peripherals Figure 1.As a result, the system designer may need to give each core exclusive access to certain resources for instance, specific peripheral devices and ensure that applications running on one core don't access resources dedicated to another core.

The presence of multiple cores can also introduce greater design complexity. For instance, to cooperate with one another, applications running on different cores may require efficient interprocess communication (IPC) mechanisms, a shared-memory data infrastructure, and appropriate synchronization primitives to protect shared resources [1].

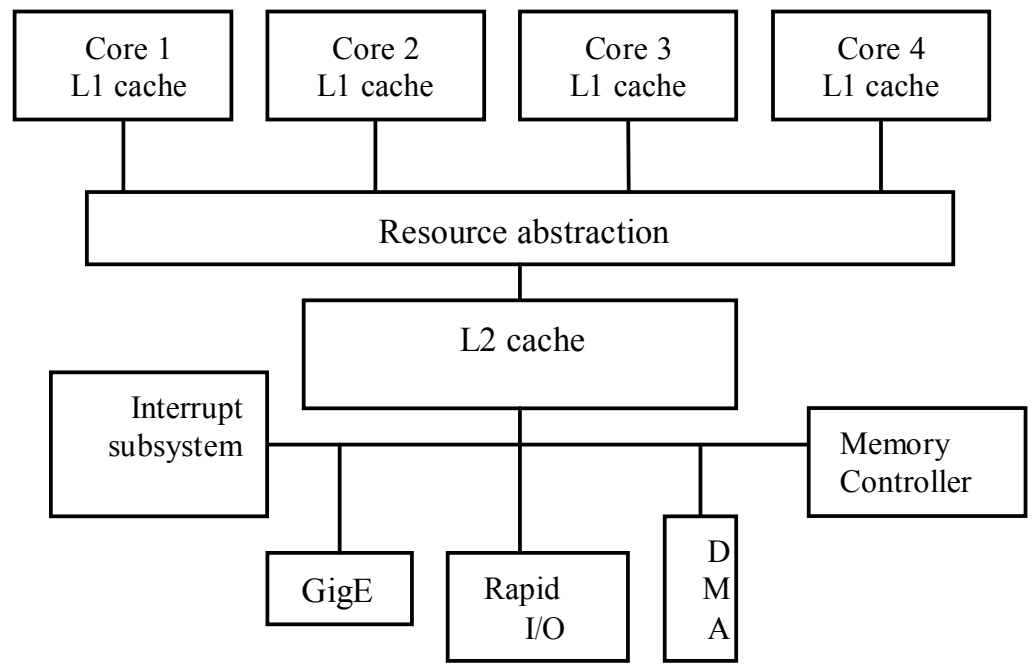

Figure. 1-Anatomy of a Multicore System.

The OS chosen for a multi-core design can significantly reduce, or increase, the effort required to address these challenges. It all depends on how the OS supports the various multiprocessing modes that a multi-core chip may offer. These modes come in three basic flavors like asymmetric, symmetric and bound, but here we are interested only in BMP (bound multiprocessing). With BMP, an application locked to one core can't leverage other cores, even if they are idle.

## 2. Related Work

Liu and Layland have presented the rate monotonic algorithm as an optimal fixed priority scheduling algorithm, and the earliest deadline first as optimal dynamic priority scheduling algorithms. Two different scheduling and methods emerged, static priority scheduling and dynamic priority scheduling. The RM algorithm is the example of the static priority scheduling algorithm and EDF is an example of dynamic scheduling algorithm [4].

### 2.1 Rate Monotonic Algorithm

This algorithm is a fixed priority scheduling algorithm which consists of assigning the highest priority to the task having the shortest period. The shorter the period, the higher the period. At any time, the scheduler chooses to execute the task with the highest priority [2] [7].

The problem with the rate monotonic algorithm is that the schedulable bound is less than 100%.The CPU utilization of task $P_i$ is computed as the ratio of worst-case computing time *Ci* to the period *Pi*. The total utilization *Un* for *n* tasks is calculated as follows [1].

n

$$U_n = \underset{i=1}{} \frac{C_i}{T_i} \qquad (1)$$

For the RM algorithm, the worst-case schedulable bound $W_n$ for n tasks is

$$W_n = n\,(2^{\,½} - 1) \qquad (2)$$

From (2) for different values of n $W_n$ is estimated which is shown in table 1.Thus a set of tasks for which total CPU utilization is less than 69% will always meet all deadlines. All tasks will be guaranteed to meet their deadlines if $U_n < W_n$.If $U_n > W_n$, then the subset of highest-priority tasks S such that $U_s < W_s$ will be guaranteed to meet all deadlines, and thus form the critical set. Another problem with RM is that it does not support dynamically changing periods, a feature required by dynamically reconfigurable systems.

Table 1
CPU utilization

| N | $W_n$ |
|---|---|
| 1 | 100% |
| 2 | 83% |
| 3 | 78% |
| | 69%(ln 2) |

### 3. Task Model

Here we assume that the real time system consists of m identical processors. Each task $T_i$ is characterized by its arrival time $_i$, ready time $R_i$, worst case computation time Ci and a dead line $D_i$.

The figure 2 below shows the processor configuration. The model is so designed that we consider a system with four cores, each core is dedicated for a set of tasks.

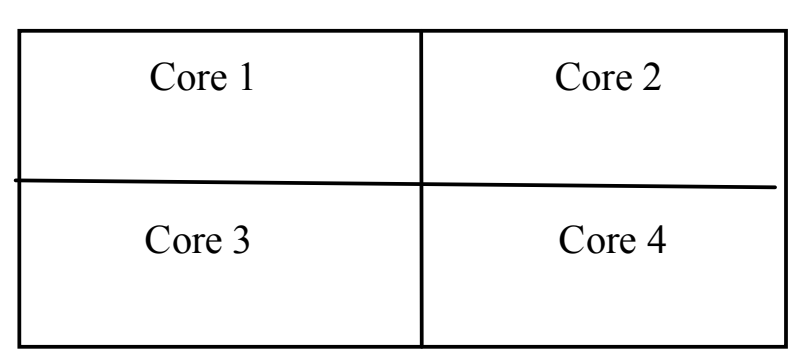

Figure.2 -Quad core processor.

Each core is represented by $\mathbf{C_i}$, ($C_1$, $C_2$, $C_3$, C4) and the tasks by Ti ($T_1$, $T_2$, $T_3 \ldots T_n$ ). Every task has set of jobs to be executed based on the EDF algorithm. These jobs are denoted by $J_{ik}$ ($J_{1,1}$,$J_{1,2}$…..Ji,n). As in the case real time processing the EDF algorithm is generally applied. Here EDF algorithm is used in scheduling of the tasks for four different cores [1].

#### 3.1 Scheduling of jobs on various processors

At the arrival of a new task the EDF immediately calculates the order of scheduling the tasks. It preempts the running task and schedules a new process according to its deadline, the preempted task is rescheduled later. It schedules the periodic tasks and other tasks based on arbitrary request, deadlines and service execution times.EDf is a dynamic algorithm schedules every instance of each incoming task according to its specific demands [5].

It reschedules periodic task for each period here the upper bound of processor utilization is 100%, if EDF has already assigned the priority for a new task, the scheduler must rearrange the priorities of other tasks until the required priority is available. In worst case, the priorities of all tasks have to be rearranged which may cause considerable overhead to the processor [8].

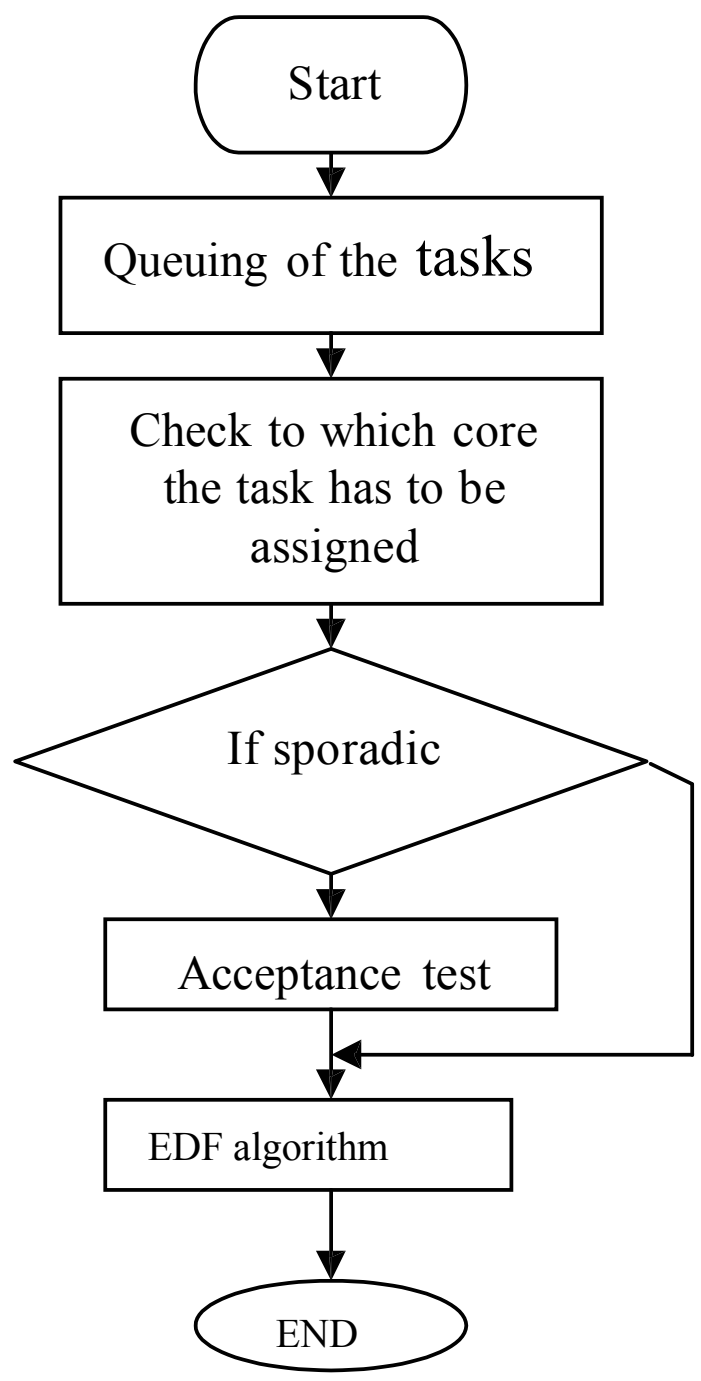

Figure 3 - Proposed method of Scheduling the task to various processors.

Lists of tasks to be executed by individual processor that is to Core1, Core 2, Core 3, and Core 4.It has been assumed that each core is dedicated to a set of tasks and there is no interdependency of the tasks associated with other cores. Here the jobs are arranged within tasks on decreasing priorities.

Here tasks are classified as periodic tasks aperiodic tasks and sporadic tasks. When sporadic tasks arise, it checks if the job can be scheduled based on the available time in the frame i.e. if a periodic task has already executed before its deadline.

## 4. Scheduling Model

Here we have assumed that task 1 has a higher priority than task 2 from time 0 until time 2.0, task 2 starts to have a higher priority at time 2.0.

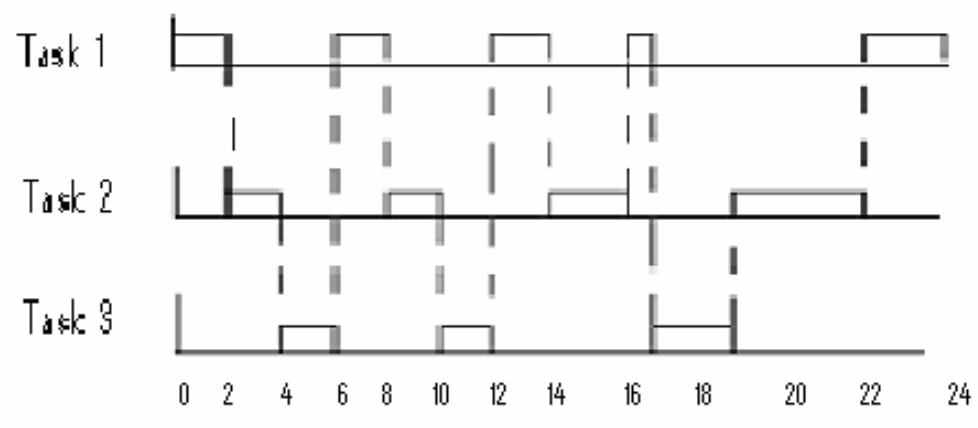

Figure 4-A dynamic EDF schedule on a uniprocessor.

At time 0, the first jobs $J_{1,1}$ and $J_{2,1}$ of both tasks are ready. Here we have assumed that $J_{1,1}$ has a higher priority and executes, $J_{2,1}$ begins executing as soon as $J_{1,1}$ releases the Core 1.J $_{3,1}$
and $J_{1,2}$ though they have the same deadline $J_{3,1}$ is executed based on the arrival time of both. Here the scheduler lowers the priorities of all the late jobs when they do not complete by their deadlines. We consider two issues one is the critical instants and the other is time delay.
In a fixed priority system where every job completes before the next job in the same task is released, a critical instant of any task Ti occurs when one of its job Ji,c is released at the same time with a job in a very higher-priority task, i.e. .$r_{i,c}$ = $r_k$, $l_k$ for some $l_k$ for every k=1,2,..,i-1.To analyze we consider a simple mathe matical expression.

$$E_i + \sum_{k=1}^{i-1} \left[ \frac{Wi,_1+wi-wk}{p_k} \right] e_k \qquad (3)$$

Where Wi denote the response time of $J_{i,1}$ and wk represents release time of the first job in $T_k$ to the instant wi+Wi when the first job $J_{i,1}$ in $T_i$ completes.

[ Wi,$_1$+wi-wk] I$p_k$ represents the jobs in $T_k$
ready for execution and ek represents the units of the processor time.

**The time demand analysis**

Here we consider a task at a time "T" starting from a task T1 with a highest priority in order of decreasing priority.

$$Wi(t)= Ei + \sum_{k=1}^{i-1} \frac{[t]}{p_k} * e_k \qquad (4)$$

## Conclusion

In this paper, we only focused on EDF for scheduling of tasks on a multicore processor. Since here the tasks are not interdependent and are executed on individual cores we have used the same technique of programming a uniprocessor.This can be implemented using various real operating systems like QNX, Vxworks.